\documentclass[prl,nofootinbib,superscriptaddress,twocolumn]{revtex4}
\usepackage[a4paper,left=1.5cm,right=1.5cm,top=3cm,bottom=3cm]{geometry}

\usepackage{amssymb,amsmath,amsfonts}
\usepackage[dvipsnames]{xcolor}
\usepackage{graphicx}
\usepackage{longtable}
\usepackage{verbatim}
\usepackage{color}
\usepackage{mdframed}
\usepackage{soul}
\usepackage{amsfonts,amssymb,mathrsfs,amsmath,esint}
\usepackage{slashed, cancel}
\usepackage{framed}
\usepackage{mdframed}
\usepackage{simplewick} 
\allowdisplaybreaks

\usepackage{latexsym}
\usepackage{graphicx}
\graphicspath{{./Figures/}}
\usepackage[dvipsnames]{xcolor}
\usepackage{booktabs}
\usepackage{datetime}
\newdateformat{mydate}{\THEDAY{ }\monthname[\THEMONTH]{ }\THEYEAR}

\usepackage{tikz}
\usepackage{color}
\usepackage{framed}
\usepackage{hyperref}
\hypersetup{colorlinks, citecolor=bluscuro, linkcolor=black, urlcolor=bluscuro}
\definecolor{rossos}{cmyk}{0,1,1,0.55}
\definecolor{bluscuro}{rgb}{0.15, 0.2, .85}
\definecolor{bluchiaro}{cmyk}{1,.3,0.,0.1}

\graphicspath{{./Figures/}}

\newcommand{\be}{\begin{equation}}
\newcommand{\ee}{\end{equation}}
\newcommand{\bea}{\begin{eqnarray}}
\newcommand{\eea}{\end{eqnarray}}
\newcommand{\beq}{\begin{equation}}
\newcommand{\eeq}{\end{equation}}
\def\beqa{\begin{eqnarray}}

\def\d{{\rm d}}

\def\eeqa{\end{eqnarray}}

\def\lsim{\mathrel{\rlap{\lower4pt\hbox{\hskip0.5pt$\sim$}}
    \raise1pt\hbox{$<$}}}         
\def\gsim{\mathrel{\rlap{\lower4pt\hbox{\hskip0.5pt$\sim$}}
    \raise1pt\hbox{$>$}}}         

\usepackage[normalem]{ulem}
\usepackage{soul}

\newcommand\ML[1]{\color{blue}#1}
\newcommand\MLL [1]{\color{orange} ML: #1}

\begin{document}

\title{Cross-correlating Astrophysical and Cosmological Gravitational Wave Backgrounds with the Cosmic Microwave Background}

\author{A. Ricciardone}
\address{Dipartimento di Fisica e Astronomia ``G. Galilei",
Universit\`a degli Studi di Padova, via Marzolo 8, I-35131 Padova, Italy}

\address{INFN, Sezione di Padova,
via Marzolo 8, I-35131 Padova, Italy}

\author{L. Valbusa Dall'Armi}
\address{Dipartimento di Fisica e Astronomia ``G. Galilei",
Universit\`a degli Studi di Padova, via Marzolo 8, I-35131 Padova, Italy}

\address{INFN, Sezione di Padova,
via Marzolo 8, I-35131 Padova, Italy}

\author{N. Bartolo}
\address{Dipartimento di Fisica e Astronomia ``G. Galilei",
Universit\`a degli Studi di Padova, via Marzolo 8, I-35131 Padova, Italy}

\address{INFN, Sezione di Padova,
via Marzolo 8, I-35131 Padova, Italy}

\author{D. Bertacca}
\address{Dipartimento di Fisica e Astronomia ``G. Galilei",
Universit\`a degli Studi di Padova, via Marzolo 8, I-35131 Padova, Italy}

\address{INFN, Sezione di Padova,
via Marzolo 8, I-35131 Padova, Italy}

\author{M. Liguori}
\address{Dipartimento di Fisica e Astronomia ``G. Galilei",
Universit\`a degli Studi di Padova, via Marzolo 8, I-35131 Padova, Italy}

\address{INFN, Sezione di Padova,
via Marzolo 8, I-35131 Padova, Italy}

\author{S. Matarrese}
\address{Dipartimento di Fisica e Astronomia ``G. Galilei",
Universit\`a degli Studi di Padova, via Marzolo 8, I-35131 Padova, Italy}

\address{INFN, Sezione di Padova,
via Marzolo 8, I-35131 Padova, Italy}

\address{INAF - Osservatorio Astronomico di Padova, Vicolo dell'Osservatorio 5, I-35122 Padova, Italy}

\address{Gran Sasso Science Institute, Viale F. Crispi 7, I-67100 L'Aquila, Italy}

\date{\today}

\begin{abstract}
\noindent
General Relativity provides us with an extremely powerful tool to extract at the same time astrophysical and cosmological information from the Stochastic Gravitational-Wave Backgrounds (SGWBs): the cross-correlation with other cosmological tracers, since their anisotropies share a common origin and the same perturbed geodesics. In this letter we explore the cross-correlation of the cosmological and astrophysical SGWBs with Cosmic Microwave Background (CMB) anisotropies, showing that future GW detectors, such as LISA or BBO, have the ability to measure such cross-correlation signals. We also present, as a new tool in this context, constrained realization maps of the SGWBs extracted from the high-resolution CMB {\it Planck} maps.  This technique allows, in the low-noise regime, to faithfully reconstruct the expected SGWB map by starting from CMB measurements.
\end{abstract}

\maketitle
\paragraph{Introduction} 

An important goal for future Gravitational Wave (GW) experiments will be that of detecting and characterizing the SGWB~\cite{Audley:2017drz, Kawamura:2006up, Evans:2016mbw, Sathyaprakash:2011bh, Maggiore:2019uih}. Such a prospect has been recently made even more tantalizing, in light of the recent claim by the NANOGrav Collaboration of a possible detection of a stochastic GW signal~\cite{Arzoumanian:2020vkk}. In nature, we expect the SGWB to be generated by two main contributions: a guaranteed one coming from late time unresolved astrophysical sources (AGWB) and another one coming from a variety of different possible physical processes in the Early Universe, such as inflation, preheating, phase transitions or topological defects (CGWB) (see e.g. \cite{Maggiore:1900zz, Regimbau:2011rp, Guzzetti:2016mkm, Caprini:2018mtu} for reviews).

The contribution of these backgrounds to the Universe energy budget is described by an average energy density parameter $\Omega_{\rm GW} (f)$ (the monopole). Forecasts of detection prospects for this monopole signal with ground and space-based detectors were shown in, e.g.,~\cite{Caprini:2015zlo, Bartolo:2016ami, Caprini:2019egz, Auclair:2019wcv, Maggiore:2019uih, Sathyaprakash:2009xt, Guo:2018npi, Luo:2015ght}. Such backgrounds are however also expected to display anisotropies (direction dependence) in the GW energy density $\Omega_{\rm GW} (f, \hat{n})$, which can be generated either at the time of their production~\cite{Bethke:2013aba, Bethke:2013vca,Ricciardone:2017kre, Geller:2018mwu, Bartolo:2019zvb, Adshead:2020bji, Malhotra:2020ket} or during their propagation in our perturbed universe~\cite{Contaldi:2016koz,Bartolo:2019oiq,Bartolo:2019yeu, Domcke:2020xmn}. While the former generation mechanism is source/model-dependent, the latter is model-independent and ubiquitous for all the SGWB sources. Tools and techniques to measure anisotropic components of the SGWB have been developed in~\cite{Giampieri:1997ie, Allen:1996gp, Cornish:2001hg, Seto:2004ji, Kudoh:2004he, Kudoh:2005as, Renzini:2018vkx, Contaldi:2020rht, Mentasti:2020yyd}. The physics that governs SGWB anisotropies (at least in the geometric optics limit) displays strong analogies with that underlying Cosmic Microwave Background (CMB) fluctuations~\cite{Dodelson:2003ft,Bartolo:2006cu,Bartolo:2006fj,DallArmi:2020dar}. Moreover, General Relativity predicts a {\em non-zero spatial correlation between the SGWB and the CMB}, since gravitons share their perturbed geodesics with CMB photons.
The main goal of this paper is that of exploring in detail and exploiting the cross-correlation of the two main SGWB sources (astrophysical and cosmological) with the CMB, focusing on effects which are mostly significant at large angular scales, where future GW interferometers will be sensitive. 
Previous cross-correlation studies between CMB and other probes, e.g.~Large-Scale Structure (LSS), have mainly focused on the correlation between the Integrated Sachs-Wolfe (ISW) effect and weak lensing~(e.g. \cite{Hu:2004yd, Hirata:2008cb,Chisari:2014xia}). 
In the case of the CGWB, we include the impact of different effects (i.e., Sachs-Wolfe (SW), ISW and Doppler) on the spectrum and we give a physical interpretation of its features at different multipoles.
Following~\cite{DallArmi:2020dar}, we modify the publicly available Boltzmann code CLASS~\cite{Lesgourgues:2011re} and we extract both the CGWB$\times$CGWB auto-correlation and the CGWB$\times$CMB angular cross-spectrum.  We then perform a Signal-to-Noise Ratio (SNR) analysis for these cross-correlation signals and find a detectable correlation between the two backgrounds. In this forecast, we consider full-sky, {\it Planck} SMICA CMB maps~\cite{planckmap} and the latest specifications of the space-based interferometers LISA~\cite{Audley:2017drz} and Big Bang Observer (BBO)~\cite{Corbin:2005ny}. 
We also explore the AGWBxCMB cross-correlation, focusing in this case on the contribution generated only by Binary Black Holes (BBHs), which is very likely to be observed by Advanced LIGO~\cite{Regimbau:2016ike}. As shown in~\cite{Cusin:2017fwz, Jenkins:2018nty, Bertacca:2019fnt}, the AGWB anisotropies are characterized by many effects: galaxy density perturbations, redshift-space distorsions, local gravitational potential, and the ISW effect. However, here we consider only the former term, since we expect it to provide the dominant contribution. 
We compute the AGWB$\times$CMB angular power spectrum taking into account all the astrophysical dependencies and LIGO latest constraints~\cite{Abbott:2020gyp,Abbott:2021xxi}. We forecast the expected SNR that will be measured by future GW detectors showing the noise level requirements to have a detection.
\vskip .2cm 
Finally, we exploit the SGWB$\times$CMB cross-correlation signal as a tool to produce constrained realization maps of the SGWB, obtained from high-resolution CMB maps from {\it Planck}. To our knowledge, this approach
is new in this specific context and it is of particular interest when considering the $\rm CMB\times CGWB$ signal, because the correlation between CMB and CGWB at low multipoles is approximately one. This means that, in the low-noise regime, the CGWB map at large angular scales is {\em univocally} determined by the CMB one. 
We generate constrained realizations also for the AGWB, which shows however a smaller degree of correlation with the CMB, compared to the CGWB. 
\noindent
\paragraph{Cross-correlation CGWB$\times$CMB} \label{Cross-correlation_section} To study the cross-correlation between the CMB and the CGWB anisotropies we solved the Boltzmann equation for the photon and graviton distribution functions $f _{\rm GW}$, $f_{\rm \gamma}$. The distribution functions for the gravitons and photons can be expanded as a leading term, homogeneous and isotropic, plus a first-order contribution rescaled in terms of the functions $\Gamma$ and $\theta$ respectively. The quantity $\Gamma$ is related to the perturbation of the GW energy density, specifically to the SGWB density contrast $\delta_{\rm CGWB}$ and to the CGWB energy density fractional contribution $\bar{\Omega}_{\rm CGWB}$~\cite{Bartolo:2019oiq, Bartolo:2019yeu}, $\delta_{\rm CGWB}  = ( 4 -  \partial \ln \, {\bar \Omega}_{\rm CGWB} \left( \eta ,\, q \right)/ \partial \ln \, q ) \, \Gamma \left( \eta ,\, \vec{x} ,\, q ,\, {\hat n} \right)$, where $q$ is the graviton comoving momentum. In this work we consider a power-law dependence of the GW energy density on frequency, $\bar{\Omega}_{\rm CGWB}\propto q^{n_T}$, so the above equation simplifies to $\delta_{\rm CGWB}=(4-n_T)\Gamma$, where $n_T$ is the tensor spectral index. Solving the Boltzmann equation at linear level around a Friedmann-Lemaitre-Robertson-Walker (FLRW) background metric which, in the Poisson gauge, $ds^2=a^2(\eta)\left[ -e^{2\Phi} d\eta^2+e^{-2\Psi}\delta_{ij}  dx^i dx^j\right]\,$, we find the following solutions for the scalar induced contribution (see~\cite{Bartolo:2019oiq, Bartolo:2019yeu} for more details on the derivation)

\begin{figure}[t!]
\centering 
\includegraphics[width=0.45\textwidth]{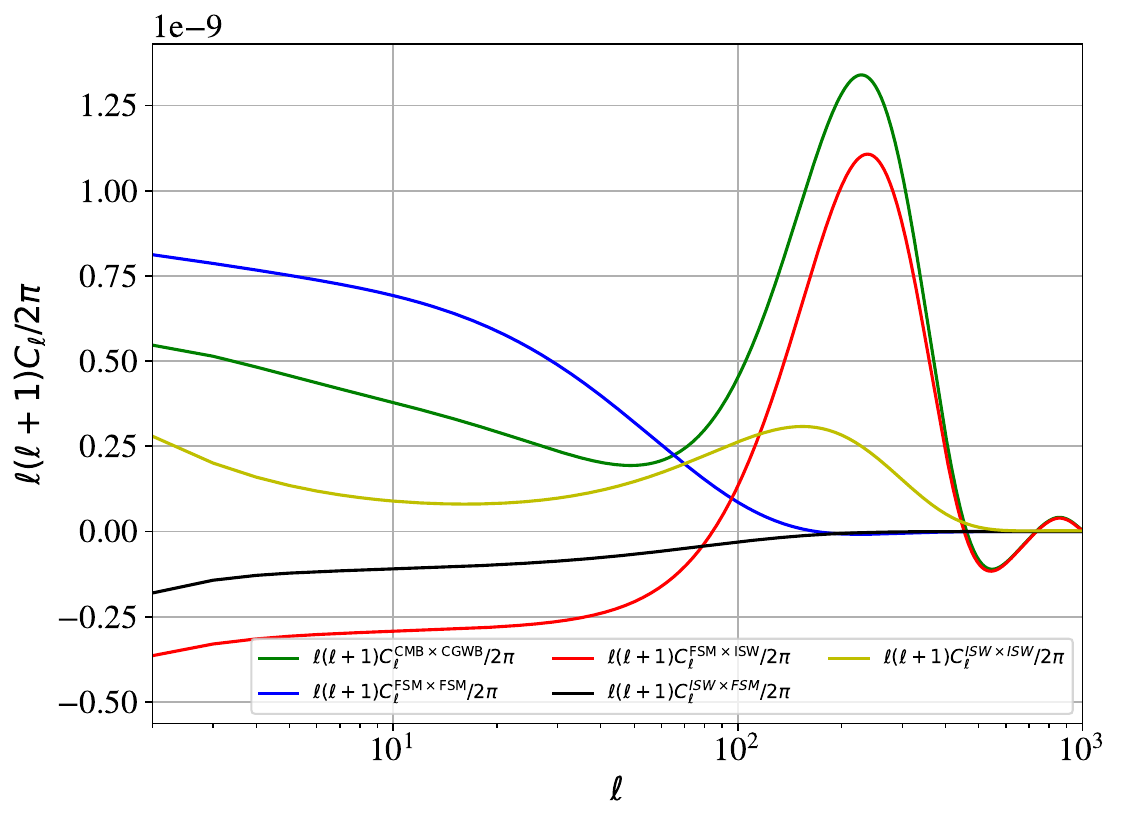}
\caption{\it Contributions to the $\rm CMB\times CGWB$ cross-correlation angular power spectrum.} 
\label{TOTAL}
\end{figure}
\begin{equation}
\begin{split}
\frac{\Gamma_{\ell m,S}(\eta_0)}{4\pi(-i)^\ell}= &\int \frac{d^3k}{(2\pi)^3}Y_{\ell  m}^*(\hat{k})\times \\ & \times \Bigl[\Bigl(\Gamma_I(\eta_i,k,q)+\Phi(\eta_i,k)\Bigl)j_\ell[k(\eta_0-\eta_i)]+\\ & +\int_{\eta_i}^{\eta_0}d\eta \Bigl(\Phi^\prime(\eta,k)+\Psi^\prime(\eta,k)\Bigl)j_\ell[k(\eta_0-\eta)]\Bigl], \\
\frac{\theta_{\ell m,S}(\eta_0)}{4\pi(-i)^\ell}= & \int \frac{d^3k}{(2\pi)^3}Y_{\ell m}^*(\hat{k})\times \\ & \times \int_{\eta_i}^{\eta_0}d\eta\Bigl[g(\eta)\Bigl(\theta_0(\eta,k)+\Phi(\eta,k)\Bigl) j_\ell[k(\eta_0-\eta)]\\& +g(\eta) \frac{-iv_b(\eta,k)}{k}\frac{d}{d\eta}j_\ell[k(\eta_0-\eta)] +\\&+ e^{-\tau(\eta)}\Bigl(\Phi^\prime(\eta,k)+\Psi^\prime(\eta,k)\Bigl)  j_\ell[k(\eta_0-\eta)]\Bigl]\, .
\end{split}
\label{alm_definition}
\end{equation}
The scalar sector for the CMB is characterized by the sum of the SW (second row in the equation for $\theta_{\ell m,S}$), Doppler (third row) and ISW (fourth row) contributions (see e.g.~\cite{Dodelson:2003ft}). On the other hand the CGWB is characterized by the sum of SW and ISW (second and third line in the equation for $\Gamma_{\ell m,S}$)~\cite{Bartolo:2019oiq,Bartolo:2019yeu}. Here, $\tau$ is the optical depth, defined as $\tau(\eta)\equiv\int_{\eta}^{\eta_0}d\eta^\prime n_{e} \sigma_T a$, with $n_e$ the free electron number density and $\sigma_T$ the Compton cross section, $v_b$ is the velocity of baryons, and $g$ is the visibility function, defined as $g(\eta)\equiv-\tau^\prime(\eta)e^{-\tau(\eta)}$~\cite{Dodelson:2003ft}. The initial condition contribution, which is model/source dependent, is taken here as $\Gamma_I(\eta_i,k,q) = -2/(4-n_T)\Phi(\eta_i,k)$ (single-clock adiabatic case) with $n_T=0$.
We expand the solutions of the Boltzmann equation in spherical harmonics, $\Gamma ( {\hat n} ) = \sum_\ell \sum_{m=-\ell}^\ell \Gamma_{\ell m} \, Y_{\ell m} ( {\hat n} )$, where ${\hat n}$ is the direction of the photon and GW trajectory. The angular auto and cross-correlation spectra are 
\begin{eqnarray}
 \langle \delta_{\ell m}^{\rm CGWB} (\eta_0)\,\delta^{*\rm CGWB}_{\ell^\prime m^\prime}(\eta_0) \rangle &\equiv &\delta_{\ell \ell^\prime}\delta_{m m^\prime} C_\ell^{\rm{CGWB}}\,, \nonumber\\
\langle \theta_{\ell m} (\eta_0)\,\theta^*_{\ell^\prime m^\prime}(\eta_0) \rangle &\equiv &\delta_{\ell \ell^\prime}\delta_{m m^\prime} C_\ell^{\rm{CMB}}\,,\nonumber \\
\langle \theta_{\ell m} (\eta_0)\,\delta^{*\rm CGWB}_{\ell^\prime m^\prime}(\eta_0) \rangle &\equiv &\delta_{\ell \ell^\prime}\delta_{m m^\prime} C_\ell^{\rm{CMB}\times \rm{CGWB}}\,. 
\label{definition_cross_correlation}
\end{eqnarray}
The cross-correlation spectrum is the sum of four terms, where each of them represents the correlation between either the freestream monopole (FSM)\footnote{The FSM is the monopole at the last scattering surface that propagates until the present epoch. The FSM of the CGWB is due to the SW effect only, while the FSM of the CMB is due both to the SW effect and to the acoustic peaks.} or the ISW, for each of the two backgrounds,
\begin{eqnarray}
C_\ell^{\rm{CMB}\times \rm{CGWB}}&=&C^{\rm{FSM}\times \rm{FSM}}_\ell+C^{\rm{FSM}\times \rm{ISW}}_\ell+C^{\rm{ISW}\times \rm{FSM}}_\ell\nonumber\\&+&C^{\rm{ISW}\times \rm{ISW}}_\ell.
\label{cross_correlations}
\end{eqnarray}
We did not include the Doppler cross-correlation since we verified that its impact on the SNR of the cross-correlation is negligible. However in Appendix A, for completeness we plotted the spectra including this contribution.
We modified the public code CLASS, originally developed for the computation of CMB anisotropies \cite{Lesgourgues:2011re}, and adapted it to the CGWB, as in~\cite{DallArmi:2020dar}, in order to study the cross-correlation angular spectrum. In Fig. \ref{TOTAL} we have depicted the numerical results of Eq. \eqref{cross_correlations}, both for the single contributions and for the total cross-correlation signal. In this computation, we have considered a scale-invariant tensor power spectrum, $n_T=0$.\\
The main features displayed by the spectra in Fig. \ref{TOTAL} can be better understood by considering the properties of the Spherical Bessel functions, which appear in Eq.~\eqref{alm_definition}. It can be shown that, when we compute the angular power spectrum, the product of the two Spherical Bessel integrated over $k$, differs from zero only if the Bessels are peaked in the same time interval, which becomes more and more narrow by increasing the multipole considered~\cite{Liguori:2003mb}. 
This means that if two anisotropies are generated at different times, the cross-correlation between them is non-vanishing only if the spatial separation of the events that generated the anisotropies is {\em much smaller} than the scale of the perturbation considered, otherwise the two events are uncorrelated. \\
For instance, we can observe that the contribution $C_\ell^{\rm FSM\times FSM}$ goes to zero for $\ell\geq 100$, even if ${C}_\ell ^{\rm CGWB}$ and $C_\ell^{\rm CMB}$ are non-vanishing. This is due to the fact that the contribution from the FSM of the CGWB is generated at the end of inflation, $\eta_i\approx 0$, while the CMB one is generated at recombination, $\eta_* \approx 280\, \rm Mpc$, so there is a suppression which goes as $C_\ell^{\rm FSM\times FSM}\propto (\eta_0-\eta_*)^{\ell}/(\eta_0-\eta_i)^{\ell}\approx 0.98^\ell$.
Even if $\eta_0-\eta_i\approx \eta_0-\eta_*$, because of the exponential dependence on $\ell$, the separation between the last scattering surface and the ``surface" of the end of inflation (which is assumed as the ``last scattering" surface for gravitons) makes the events uncorrelated at small scales. In a similar way, we can explain the peak of the spectrum at $\ell\approx 300$ as a resonance between the early-ISW of the CGWB and the first acoustic peak of the CMB, because both effects occurred at $\eta\approx \eta_*$, around this multipole.
On the other hand, at larger multipoles, the cross-correlation goes to zero, because the CMB and the CGWB anisotropies, which are generated at different epochs, for the scales considered, are not correlated. It is also interesting to notice that the cross-correlations $C_\ell^{\rm FSM\times ISW}$ and $C_\ell^{\rm ISW \times FSM}$, depicted in Figure \ref{TOTAL}, are negative on large scale since the transfer functions of the FSM and of the ISW have opposite sign. Intuitively, the SW of the CGWB represents the energy {\em lost} by a graviton which escapes from a potential well, while the ISW of the CMB corresponds to the energy {\em gained} from the decay of scalar perturbations when a photon is crossing them. This means that the metric perturbations give rise to effects of opposite sign. The cross-correlation is negative because when the energy of one background increases the energy of the other one decreases.

\begin{figure}[t!]
\centering
\includegraphics[scale=0.48]{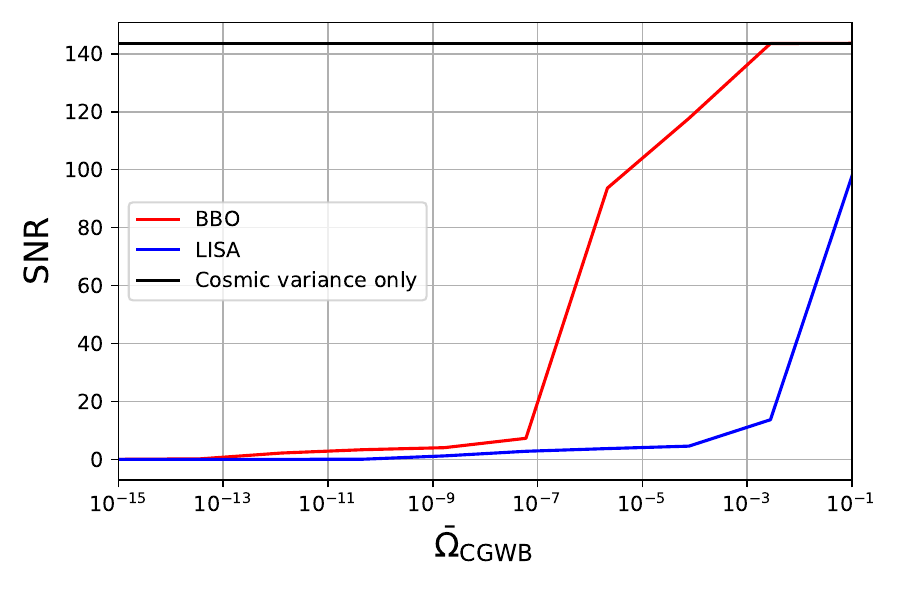}
\vskip -.5cm
\caption{\it SNR of the angular power spectrum of the cross-correlation versus monopole energy density of the CGWB, in the case of BBO (blue line), LISA (red line), and noiseless (black line) cases.}
\label{snr_monopole}
\end{figure}

\paragraph{SNR computation}
Having computed the expected CGWB$\times$CMB signal, we can now study its detectability by performing a SNR analysis. The expression for the SNR is\footnote{In our analysis we consider an ideal case assuming that both the CGWB and the CMB maps used to calculate the cross-correlation are full sky and are completely free of contamination (i.e., $f_{\rm sky}=1$). In a more realistic case we would need to consider foreground contamination in both backgrounds. This would lead to partial sky coverage -- i.e., a reduction of the SNR by a factor $f_{\rm sky} < 1$ -- and it would require the incorporation of component separation uncertainties, with a further signal degradation effect.}
\begin{equation}
{\rm SNR}^2=\sum_{\ell=2}^{\ell_{\rm max}}\frac{\left(C_\ell^{\rm CMB \times CGWB}\right)^2}{\sigma_\ell^2}\, ,
\end{equation}
where the angular spectrum of the noise $N_\ell$ is given by the sum of instrumental noise and cosmic variance,
\begin{equation}
\sigma^2_\ell=\frac{\left(C_\ell^{\rm CMB \times CGWB}\right)^2+\left(C_\ell^{\rm CGWB}+N_\ell^{\rm CGWB}\right)C_\ell^{\rm CMB}}{2\ell+1}\,.
\label{equation_noise}
\end{equation}
In our analysis we have neglected the instrumental noise of the CMB experiment, since it is well known that at the multipoles we are interested in ($\ell \lesssim 100 $), available CMB data are completely cosmic variance dominated.
\\
On the interferometer side, we have instead considered the expected noise levels of LISA and BBO, following~\cite{Mukherjee:2019oma,Alonso:2020rar, Malhotra:2020ket}; for BBO, we have only considered the two units of aligned detectors in the star-of-David~\cite{Yagi:2011wg,Yagi:2013du}. The corresponding noise angular power-spectra are reported in the supplemental material. We can notice that the low resolution of space-based detectors limits in practice the SNR analysis to the first few multipoles (ways on how to improve the angular resolution of GW detectors are discussed in~\cite{Baker:2019ync}).  \\
We have estimated the SNR as a function of the GW background energy density $\bar{\Omega}_{\rm CGWB}$. We considered either an ideal, noiseless case scenario,  or noise levels expected by e.g. LISA or BBO. As we can see from Fig.~\ref{snr_monopole} for the LISA detector, if the amplitude of the monopole GW signal is large enough ($\bar{\Omega}_{\rm CGWB}\approx 10^{-8}$), the SNR for the cross-correlation can be of order unity. In the BBO case, due the the better sensitivity, a lower GW signal is instead sufficient ($\bar{\Omega}_{\rm CGWB}\approx  10^{-12}-10^{-11}$) to produce a detectable signal (SNR $\sim2-5$). Such GW monopole amplitudes can be generated by many primordial mechanisms characterized by a blue (i.e., $n_T>0$) tensor power spectrum, like axion-inflation models~\cite{Cook:2011hg,Namba:2015gja, Domcke:2016bkh, Dimastrogiovanni:2016fuu}, models of solid and supersolid inflation~\cite{Endlich:2012pz, Bartolo:2015qvr, Ricciardone:2016lym} and other post-inflationary mechanisms as phase-transition~\cite{Caprini:2015zlo, Caprini:2019egz}, cosmic strings~\cite{Auclair:2019wcv} or PBH~\cite{Garcia-Bellido:2016dkw,Domcke:2017fix,Bartolo:2018evs,Bartolo:2018rku}. Besides specifically considering LISA or BBO, it is also interesting to envisage more advanced future scenarios, in which the instrumental noise level is reduced. This could be achieved for instance by considering all the LISA TDI channels, even in some particular new combinations that suppress the detector noise~\cite{Muratore:2020mdf}, or when more space-based detectors are correlated among them (considering for instance a network LISA-Taiji~\cite{Orlando:2020oko} or LISA-BBO). Finally, if we consider an ideal, noiseless, case we can reach a large Signal-to-Noise, SNR $\sim 100$.
\paragraph{Constrained realizations}
Since the correlation coefficient between the CMB and the CGWB approaches $1$ on large scales, a natural -- yet in this context still unexplored -- approach is that of exploiting the observed CMB temperature signal in order to build constrained realizations of the expected CGWB anisotropy field.

\begin{figure*}[t!]
    \centering
    \includegraphics[width = 0.28 \textwidth]{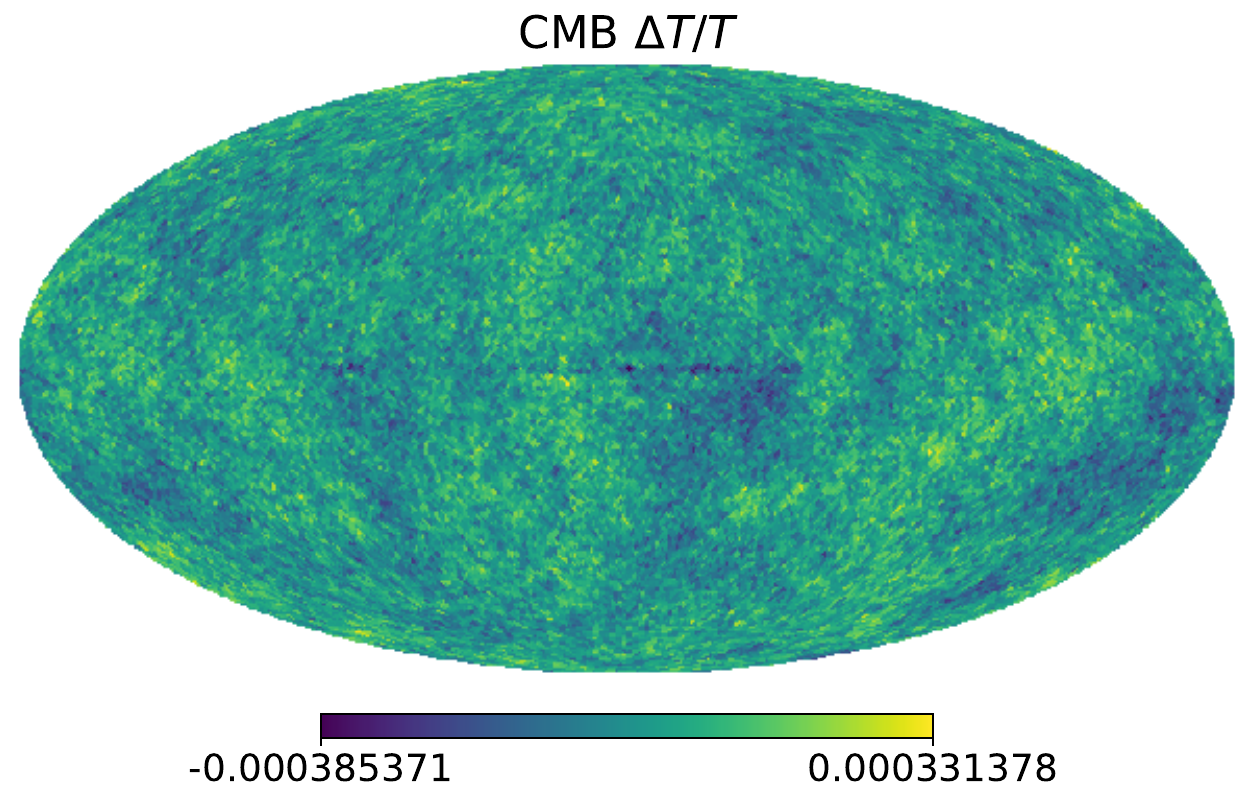}
     \includegraphics[width = 0.28 \textwidth]{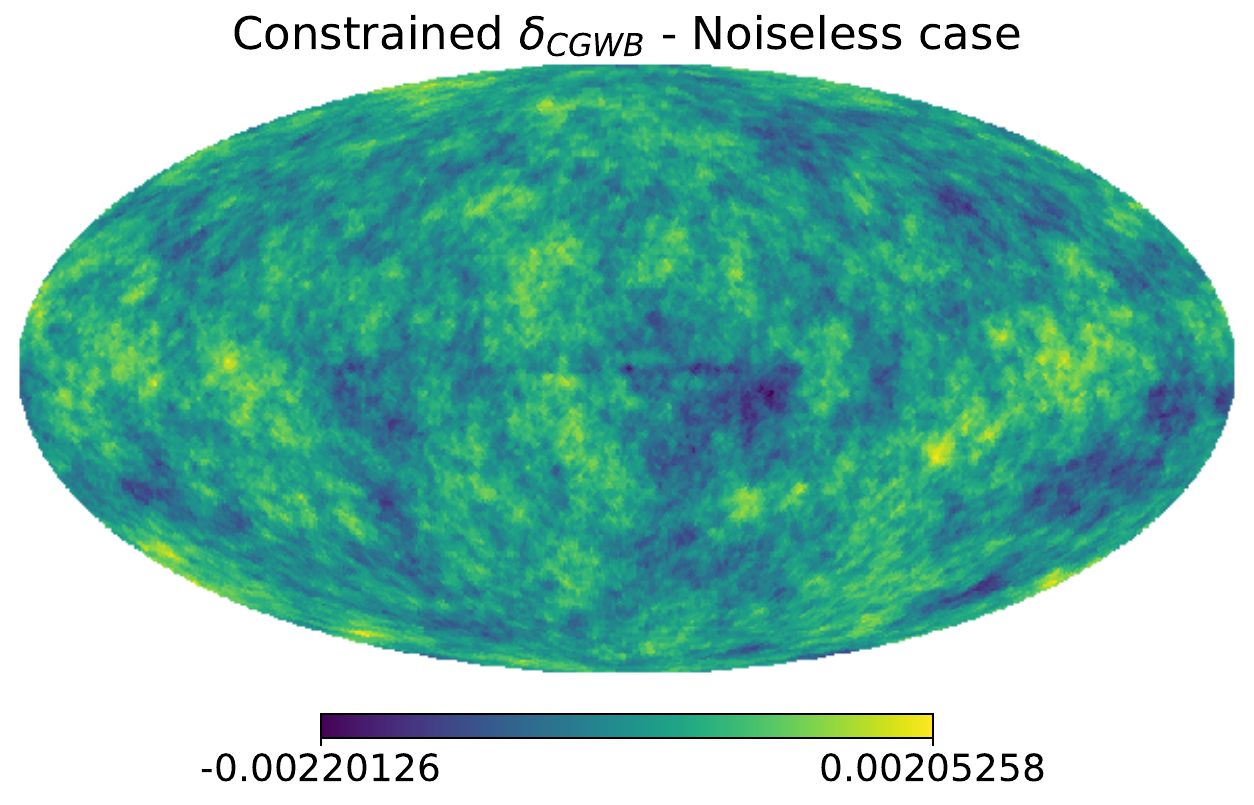}        
    \includegraphics[width = 0.28 \textwidth]{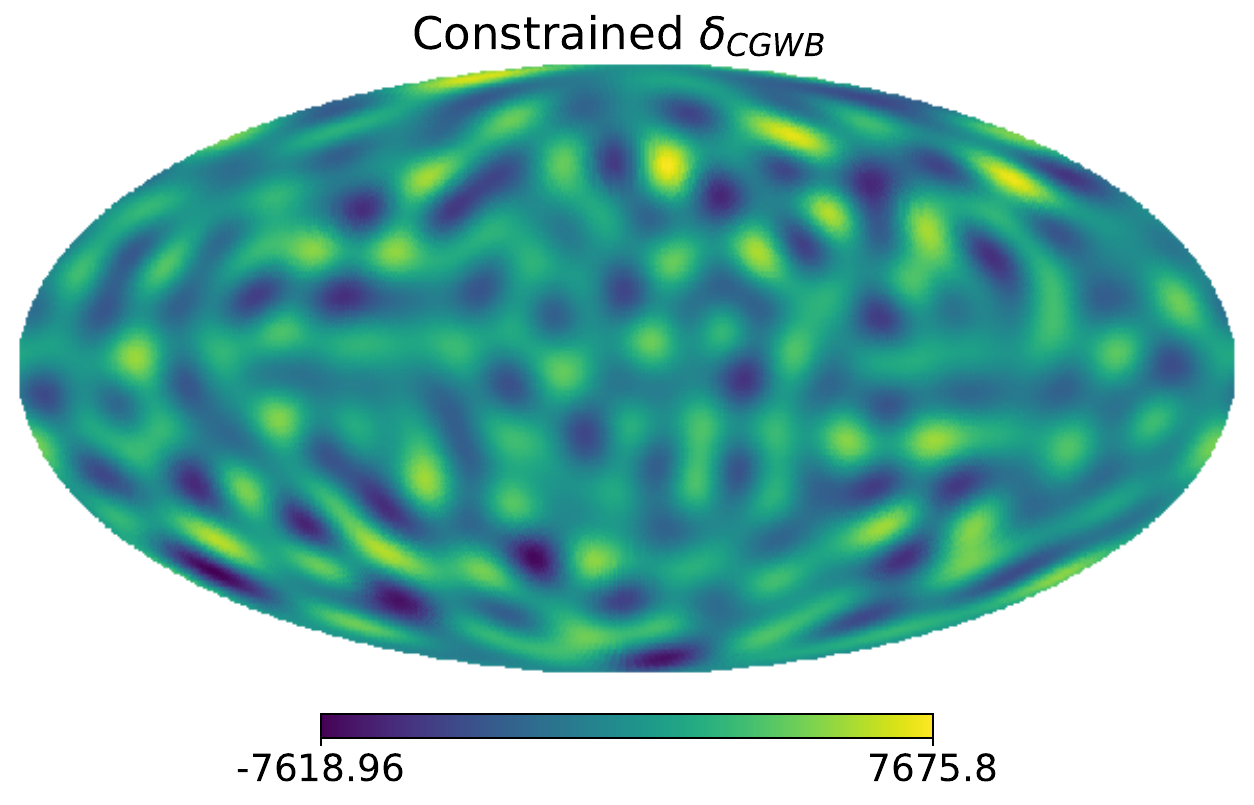} 
    \caption{{\it Left plot: Planck CMB SMICA map with $\ell_{\rm max}=200$; Central plot: constrained CGWB map with $\ell_{\rm max}=200$ in the noiseless case; Right plot: constrained CGWB map with $\ell_{\rm max}=20$. In all the maps we have fixed $N_{\rm side}=64$. }
    }
    \label{figure_realizations}
\end{figure*}

\begin{figure*}[t!]
    \centering
    \includegraphics[width = 0.28 \textwidth]{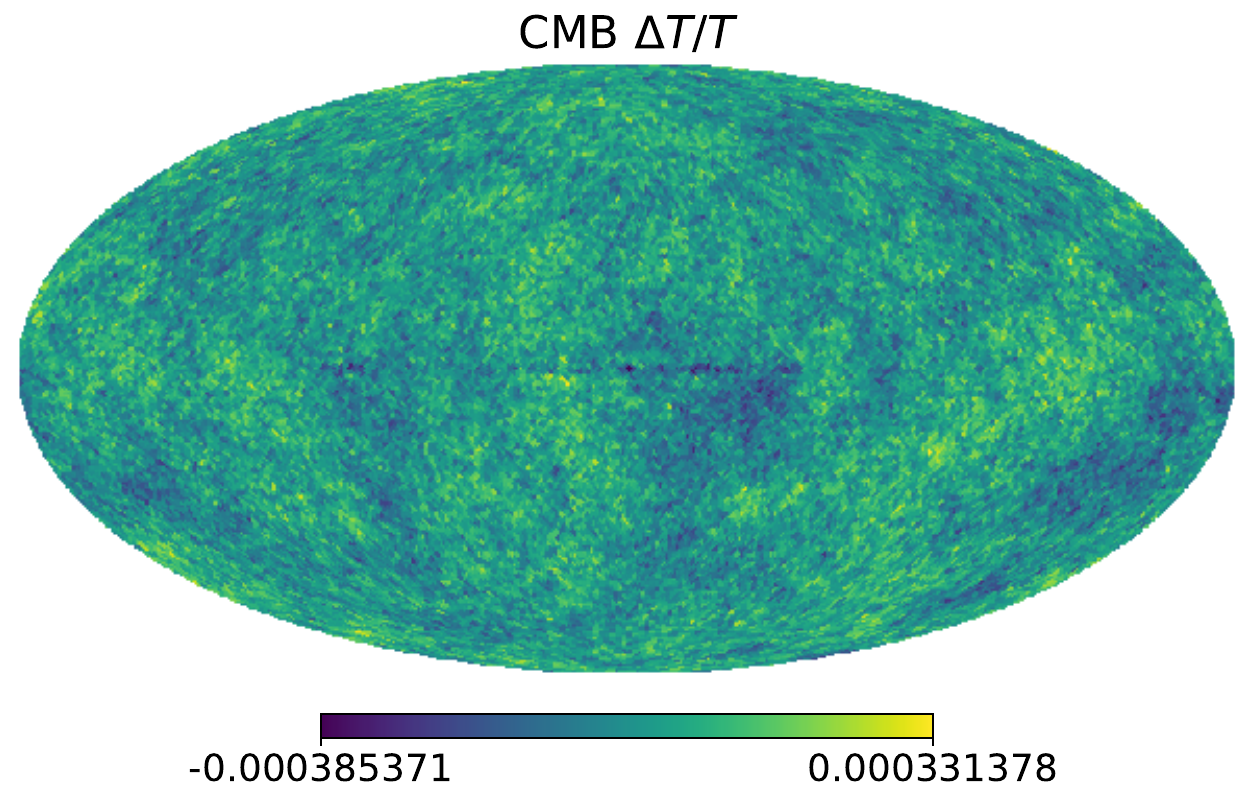}
     \includegraphics[width = 0.28 \textwidth]{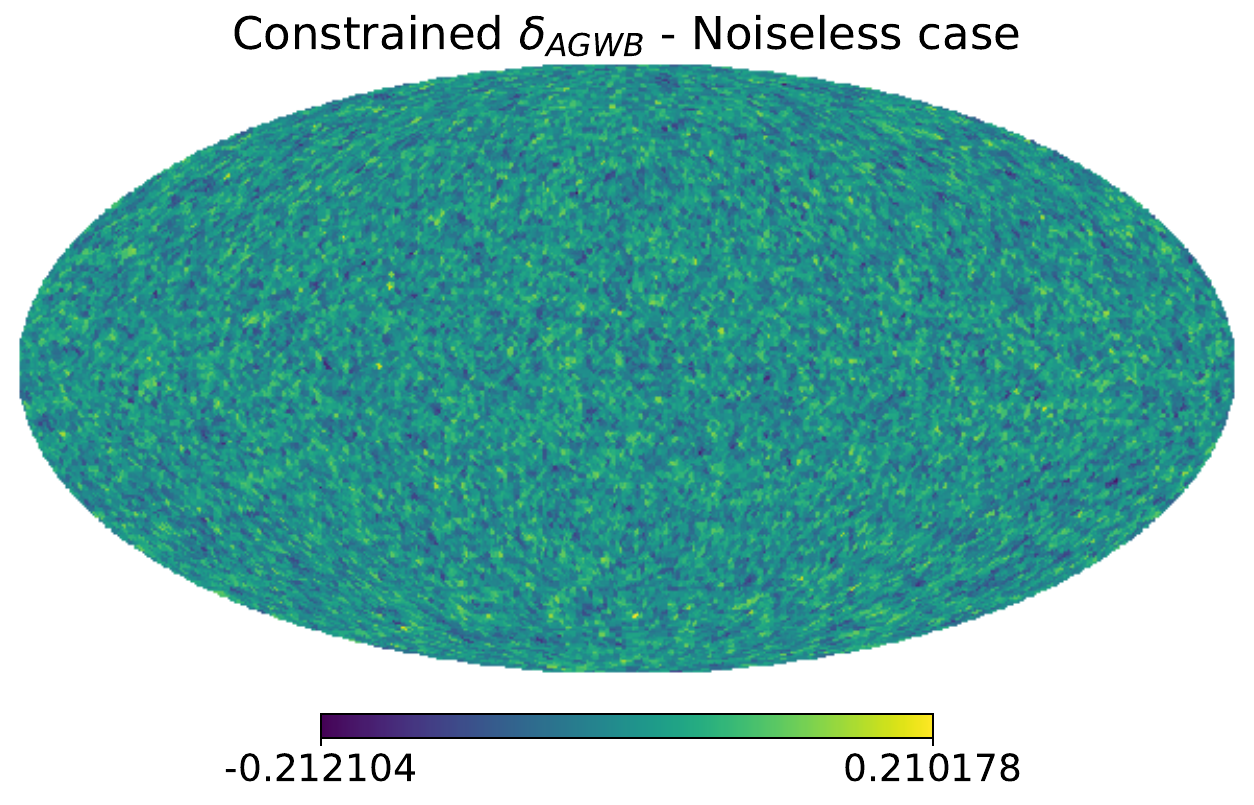}       \includegraphics[width = 0.28 \textwidth]{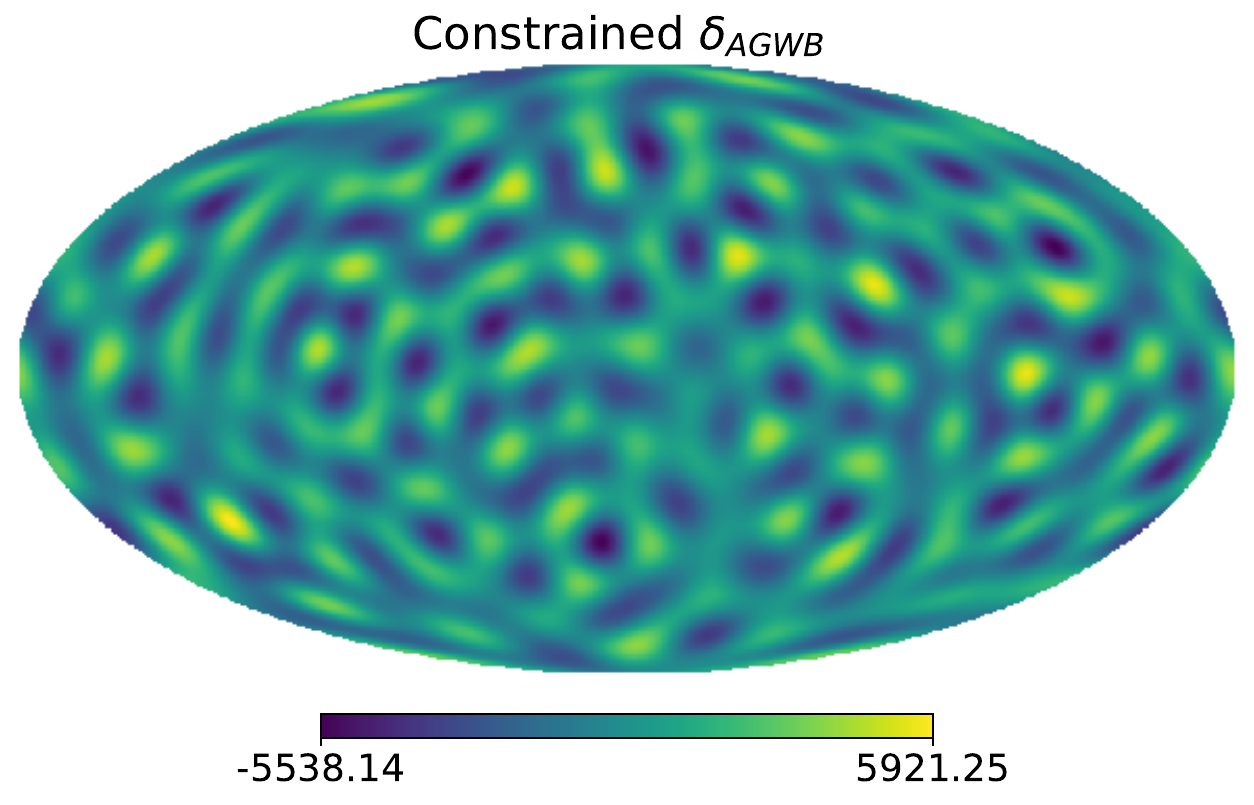} 
    \caption{\it{Left plot: Planck CMB SMICA map with $\ell_{\rm max}=200$; Central plot: constrained AGWB map with $\ell_{\rm max}=200$ in the noiseless case; Right plot: constrained AGWB map with $\ell_{\rm max}=20$. In all the maps we have fixed $N_{\rm side}=64$}. 
    }
    \label{figure_realizations_asgwb}
\end{figure*}

Let us start by briefly reviewing, in the context of our application, the general method to generate constrained realizations of Gaussian fields ~\cite{Bertschinger1987,Hoffman1991,Bucher:2011nf, Kim:2012iq, Manzotti:2014kta}.
The $a_{\ell m}$ and $\Gamma_{\ell m}$ spherical harmonics coefficients are distributed as Gaussian random variables with zero average and covariance given by a block-diagonal matrix, with $\ell_{\rm max}-2$ blocks. Each block is equal to 
\begin{equation}
C_\ell^{\rm block}=\begin{pmatrix}
C_\ell^{\rm CGWB} & C_\ell^{\rm CMB\times CGWB} \\
C_\ell^{\rm CMB \times CGWB} & C_\ell^{\rm CMB}
\end{pmatrix}.
\end{equation} 
The conditional probability of $\Gamma_{\ell m}$ given the $a_{\ell m}$ is still a Gaussian with mean and elements of the covariance matrix given by
\begin{equation}
\mu_{\ell m}=\frac{C_\ell^{\rm CMB\times CGWB}}{C_\ell^{\rm CMB}} a_{\ell m}\, ,\;\; \Sigma_{\ell m}=C_\ell^{\rm CGWB}-\frac{\left(C_\ell^{\rm CMB\times GW}\right)^2}{C_{\ell}^{\rm CMB}}\,.
\label{new_mean_equation}
\end{equation}
We extract the $a_{\ell m}$  from the full-sky, {\it Planck} SMICA CMB map using Healpix~\cite{Gorski:2004by}.  Then we generate the constrained CGWB map with mean $\mu_{\ell m}$ and covariance $\Sigma_{\ell m}$ reported above, with angular power spectra obtained considering the most recent {\it Planck} best-fit parameters~\cite{Aghanim:2018eyx}. The result of this procedure is reported in Figure \ref{figure_realizations}, where we have generated both a low resolution ($\ell_{\rm max}= 20$) and a higher resolution ($\ell_{\rm max} = 200$) constrained map. 
As stressed at the beginning of this section, until $\ell \simeq 30$, the correlation coefficient $r$ between the CMB and the CGWB, defined as $r = C_\ell^{\rm CMB\times CGWB}/\sqrt{C_\ell^{\rm CMB}C_\ell^{\rm CGWB}}$, is almost one. In the low-noise regime, the CGWB signal on large scales -- i.e., those at which interferometers are sensitive -- is then nearly deterministically predicted by the observed CMB fluctuations, considering small uncertainties on cosmological parameters (technically, this reflects in the fact that, when $r \rightarrow 1$, the variance of the constrained multipoles goes to $0$ in the cosmic variance dominated limit). For this reason, these constrained realizations of the CGWB could be used as a useful tool in the future, for example to test foreground or other systematic contamination in the data, by comparing the observed map with the CMB-based prediction. 
\paragraph{Cross-correlation AGWB$\times$CMB}
In the case of the cross-correlation between CMB and AGWB anisotropies we focused on the background contribution generated from the merging of BBHs, which is expected to be detected in the next LIGO run (O5)~\cite{Regimbau:2011rp,Zhu:2011bd}.
Using the formalism developed in~\cite{Bertacca:2019fnt}, we can write the energy density of the AGWB monopole as $\bar{\Omega}_{\rm AGWB}/4\pi = \frac{f_o}{\rho_{\rm c}}\int \d \bar{\chi}\,\d \vec{\theta}\,\frac{N(z,f_{\rm e},\vec{\theta})}{(1+z)}w(z) \, $
where $f_o$ is the measured frequency, $f_e$ is the frequency at emission, $\rho_{\rm c}$ is the critical energy density of the Universe, $\bar{\chi}$ is the comoving distance (i.e., conformal time) in the observer's frame~\cite{Schmidt:2012ne, Bertacca:2017vod}, $\vec{\theta}$ contains all the astrophysical parameters (e.g., the mass of the halo $M_{\rm h}$, the mass of the star that originated the binary $M_{\star}$, the mass of the compact BHs $\vec{m}$, the spin of the BBH, the orbital parameters and the star formation rate (SFR)) and $N(z,f_{\rm e},\vec{\theta})$ is the total comoving density of GWs reported in Appendix B, together with the details on the generation of our AGWB signal. Starting from the background we can defined the AGWB over-density as~\cite{Bertacca:2019fnt}
\begin{equation}
\begin{split}
\frac{\delta_{\rm AGWB,\ell m} }{4\pi(-i)^\ell}=\int \frac{d^3k}{(2\pi)^3}Y_{\ell m}^*(\hat{k})\int d\eta \, &\tilde{\mathcal{W}}(\eta)b\delta_m(\eta,k)\\
& \hspace{-1em}\times j_\ell[k(\eta_0-\eta)]\, ,
\end{split}
\label{eq_delta_omega_asgwb}
\end{equation}
where $b$ is the galaxy bias and $\tilde{\mathcal{W}}$ is the weight function reported in Appendix B. Since we are interested here just in an order-of-magnitude estimate of the expected correlation signal at small redshift, we have taken $b=1$, for the sake of simplicity.
\\
The cross-correlation of the AGWB with the CMB is now induced by late-ISW of the latter. In the noiseless case the SNR of the cross-correlation is $\simeq 5$  for a monopole energy density $\bar{\Omega}_{\rm AGWB}\simeq 10^{-10}$ at $f=25\, \rm Hz$. When the noise of the ground-based detectors is included, a detectable level of the cross-correlation signal requires a large amplitude of the monopole energy density, of the order $10^{-7}$, which is ruled out by latest LIGO constrains~\cite{Abbott:2021xxi}. However, the SNR will increase significantly with future space-based detectors, like LISA or BBO. In this case, a noise level $N_{\ell} \approx 10^{-26} $ (the expected noise level of LISA) allows for a possible detection (i.e., SNR  $\sim 1$) even for a monopole amplitude $\bar {\Omega}_{\rm AGWB} \approx 10^{-10}$. This is just a back of the envelope result. A more detailed analysis for the LISA detector will be performed in a separate project, taking into account the appropriate astrophysical population dependencies and evolution in the LISA band, and including all the LISA TDI channels in suitable combinations~\cite{Muratore:2020mdf}. 
We finally extract the constrained realization maps of the AGWB starting from the CMB SMICA {\it Planck} map. Also in this case we can clearly recognize in the noiseless map the seeds of the CMB anisotropies. In this case we see more differences between the two maps, because the correlation comes from the late-ISW effect, which is only a small contribution in the full CMB map. If we would consider the Advanced LIGO noise, the constrained map would be highly dominated by the noise of the interferometer, which would greatly reduce the resolution. The maps are reported in Fig. \ref{figure_realizations_asgwb}.
\paragraph{Conclusions} 
In this letter we have shown that the CMB correlates with both the AGWB and the CGWB. In the cosmological case at small multipoles the leading term is due to the cross-correlation between the two FSM of the two tracers, while at larger multipoles the spectrum is more sensitive to the cross-correlation of the early-ISW of the CGWB and of the FSM of the CGWB.
For the AGWB, we have considered as a first estimate only the galaxy density contribution with a constant bias. In this case, the correlation is induced by the late-ISW of the CMB with the density term of the AGWB. We have performed an SNR analysis to estimate the detectability of these effects for future GW interferometers, such as Advanced LIGO, LISA and BBO. 
Finally, we have obtained constrained realization maps of the CGWB and the AGWB, starting from the high-resolution {\it Planck} CMB map. 
For the CGWB map, the CMB seeds are more visible because the two are nearly perfectly correlated on large scales. This nearly deterministic prediction of the CGWB from CMB observations, in the low-noise limit, can be exploited as a useful diagnostic tool in future analyses of CGWB anisotropies.  While the noise level of forthcoming experiments is still fairly high for a meaningful application of this technique, significant improvements are expected in the future. A way to significantly lower the noise level (by orders of magnitude) would also be to combine several interferometers; we have not explicitly accounted for this possibility in our SNR forecasts.
The AGWB $\times$ CMB correlation is lower than the CGWB $\times$ CMB one. Therefore CMB-based constrained realizations are less powerful in this case. However, the power of constrained realization techniques could in this case be significantly enhanced by accounting for correlations between the AGWB and LSS observables~\cite{Canas-Herrera:2019npr, Mukherjee:2019wcg, Mukherjee:2020hyn, Mukherjee:2020mha}. We will explore this in a future work \cite{supmat2}.

Right after completing this work, we came across \cite{braglia} where the cross-correlation CGWBxCMB has been used to assess the capability of future gravitational wave interferometers to constrain Early Universe extensions to the $\Lambda$CDM model through a Fisher analysis.

\paragraph{Acknowledgments.}
\noindent
We thank A. Lewis for useful comments and discussions. 
N.B., D.B., M.L. and S.M. acknowledge partial financial support by ASI Grant No. 2016-24-H.0. M.L. was supported by the project "Combining Cosmic Microwave Background and Large Scale Structure data: an Integrated Approach for Addressing Fundamental Questions in Cosmology", funded by the MIUR Progetti di Ricerca di Rilevante Interesse Nazionale (PRIN) Bando 2017 - grant 2017YJYZAH. A.R.~acknowledges funding from MIUR through the ``Dipartimenti di
eccellenza'' project Science of the Universe.
\vskip 1.5cm

\section{Appendix A}
\begin{figure}[h!]
\centering
\includegraphics[scale=0.5]{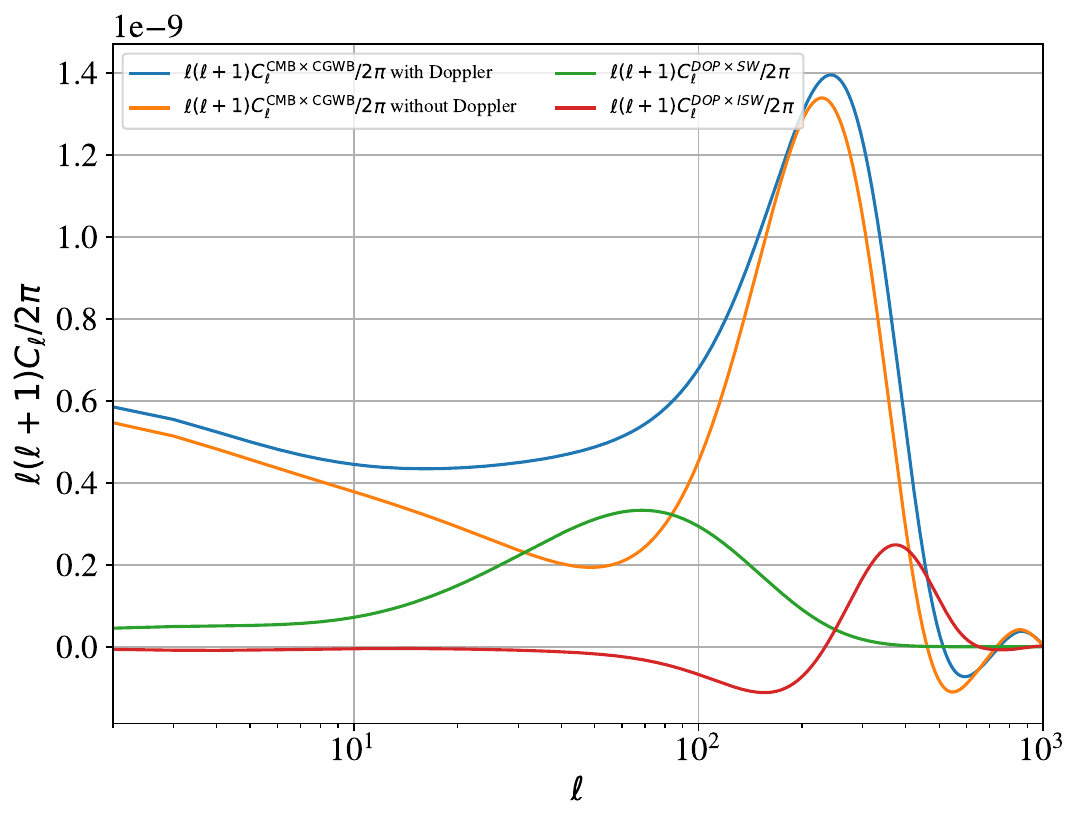}
\vskip -0.4cm
\caption{\it Doppler contributions to the $\rm CMB \times CGWB$ cross-correlation angular power spectrum.}
\label{dop_figure}
\end{figure}

As anticipated in the main text, in the cross-correlation signal $\rm CMB\times CGWB$, the Doppler effect of the former must be accounted for. Since we realize that the inclusion of this effect give a negligible contribution on the SNR, we did not include it. However, for completeness, we plot the angular cross-spectrum including also the Doppler contribution in Fig. \ref{dop_figure}. The correlation of the Doppler of the CMB with the SW and the ISW of the CGWB is non-negligible because of projection effects. Being the Spherical Bessel function of order $\ell$, $j_\ell(x)$, and its derivative, $j_\ell^\prime(x)=j_{\ell-1}(x)-(\ell+1)/xj_\ell(x)$, out of phase, if the Doppler and the SW are computed at the same time (in the CMB case at recombination), then they sum incoherently and the cross-term is zero. In the $\rm CMB \times CGWB$ case, the distance between the last scattering surface and the surface of the end of inflation, $\eta_*-\eta_i$, shifts the FSM of the CGWB, increasing the coherence between the SW of the CGWB and the Doppler of the CMB.
 At low multipoles, which are the ones that contribute most to the SNR, the Doppler terms are negligible. This justifies the choice we have done in section in the main text. The $\rm DOP \times FSM$ term has a peak around $\ell\approx 70$ due to the peak of the CMB dipole transfer function and the phase-shift $(\eta_0-\eta_*)/(\eta_0-\eta_i)$ that adds coherence between the Spherical Bessel function of the CGWB and the derivative of the Spherical Bessel function of the CMB. The peak we see in the $\rm DOP \times ISW$ spectrum is due to a resonance between the second peak of the CMB Doppler transfer function with the early-ISW of the CGWB.

\section{Appendix B}
In this appendix we report the noise angular power spectrum  used in Eq. \eqref{equation_noise} for computing the SNR. We have evaluated the angular power spectrum of the two most promising candidates to detect the CGWB anistropies, LISA and BBO. We have used the Schnell code~\cite{Alonso:2020rar}, slightly modifying it to include also BBO. 
\begin{figure}[h!]
\centering
\includegraphics[scale=0.5]{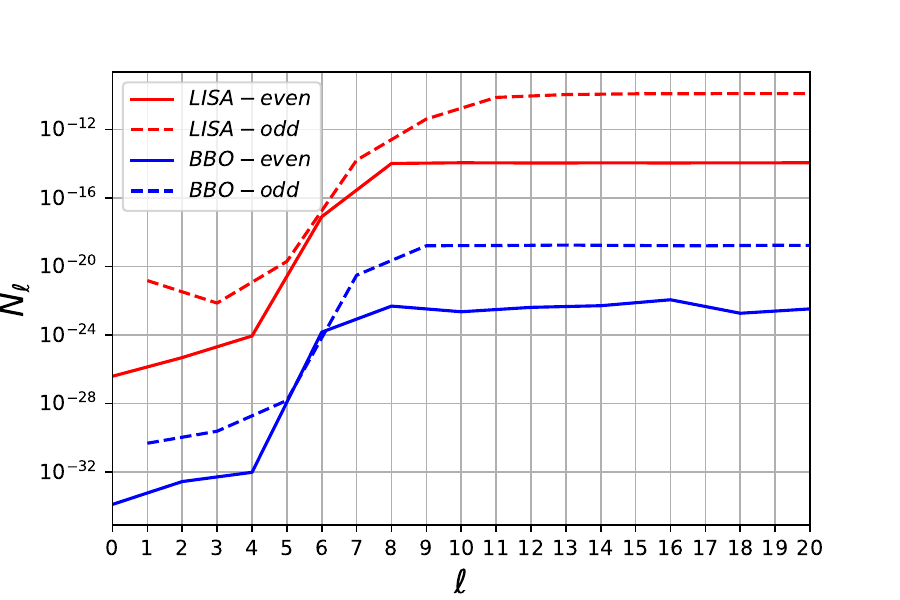}
\vskip -0.4cm
\caption{\it Noise angular power spectrum for LISA and BBO (star configuration). The relation between $N_\ell$ in the figure and $N_\ell^{\rm CGWB}$ in Eq. \ref{equation_noise} is $N_\ell^{\rm CGWB}=N_\ell/\left(\bar{\Omega}_{\rm CGWB}\right)^2$, where $\bar{\Omega}_{\rm CGWB}$ is the energy density of the gravitational waves in the case of a scale-invariant power spectrum.}
\label{nl_figure}
\end{figure}
For each detector, we have separated the noise angular power spectrum for odd and even multipoles, to study the sensitivity to different multipoles and to understand which of them give a major contribution to the SNR. \\
As expected, the lowest multipoles have the lowest $N_\ell$, leading to the dominant contribution to the SNR. To be sensitive to higher multipoles too, we require a monopole of the background large enough, in order to reduce the noise which appears in the SNR, $N_\ell/\left(\bar{\Omega}_{\rm CGWB}\right)^2$. This is in agreement with the plot of Figure \ref{snr_monopole}.

\section{Appendix C}

The weight function introduced in Eq. \eqref{eq_delta_omega_asgwb} is defined as 
\begin{equation}
\tilde{\mathcal{W}}[\eta(z)]\equiv \frac{f_{\rm o}}{\rho_{\rm c}} \frac{4\pi}{\bar{\Omega}_{\rm AGWB}} \frac{w(z) N(z,f_{\rm e},\vec{\theta})}{(1+z)}\, ,
\end{equation}
where the total comoving density of GWs, $N(z,f_{\rm e},\vec{\theta})$, can be written as
\begin{equation}
N(z,f_{\rm e},\vec{\theta}) = R(z,\vec{\theta})\frac{dE_{\rm GW,e}(f_o,z,\vec{\theta})}{df_{\rm e}d\Omega_{\rm e}}\, ,
\end{equation}
where the two terms on the RHS are the intrinsic comoving merger rate of BBHs and the energy of the GWs emitted with a frequency $f_{\rm e}$. In our analysis, we have taken into account all the stages of evolution of a binary system, namely inspiral, merger, and ringdown. For the energy spectrum of the GWs in these stages we have used the waveforms given in~\cite{Ajith:2012mn,Ajith:2007kx,Ajith:2009bn}. We have computed the merger rate of compact objects assuming a linear dependence on the star formation rate (SFR), using $R(z,\vec{\theta})=A\, {\rm SFR}(z,\vec{\theta})$, where $A$ is a factor that encodes information about the fraction of compact objects that have a companion (thus that form a binary) and the fraction of binaries that merge within an Hubble time. We have determined this constant from the latest constraints by the Advanced LIGO's and Advanced Virgo's Third Observing Run, $R(0)=19.1\, \rm Gpc^{-3}yr^{-1}$~\cite{Abbott:2021xxi}. For simplicity, here we have used the parametrization of the cosmic SFR given in~\cite{Madau:2014bja}, neglecting also contributions related to the time delay between the formation of the binary and its merge, leaving them for a more dedicated future analysis. \\
In the computation of the monopole energy density for the AGWB, we have also included a window function $w(z)$, which represents the efficiency of the detector involved in measuring the AGWB. It can be computed by integrating the PDF of the SNR from 0 until the threshold for having a GW detection of a resolved source, which we choose equal to 8~\cite{Karnesis:2021tsh},
\begin{equation}
w(z)=\int_{0}^8 \d {\rm SNR}\,  p({\rm SNR}|z,f_{\rm o},\vec{\theta})\, . 
\end{equation}

\bibliographystyle{ieeetr}
\bibliography{Biblio.bib}

\end{document}